# Deterministic positioning of colloidal quantum dots on silicon nitride nanobeam cavities

Yueyang Chen[1], Albert Ryou[1], Max R. Friedfeld[3], Taylor Fryett[1], James Whitehead[1], Brandi M. Cossairt[3], Arka Majumdar[1,2]

[1] Electrical Engineering, University of Washington, Seattle, WA 98189, USA

[2] Department of Physics, University of Washington, Seattle, WA 98189, USA

[3] Department of Chemistry, University of Washington, Seattle, WA 98189, USA

**Abstract:**

Engineering an array of precisely located cavity-coupled active media poses a major experimental challenge in the field of hybrid integrated photonics. We deterministically position solution processed colloidal quantum dots (QDs) on high quality-factor silicon nitride nanobeam cavities and demonstrate light-matter coupling. By lithographically defining a window on top of an encapsulated cavity that is cladded in a polymer resist, and spin coating QD solution, we can precisely control the placement of the QDs, which subsequently couple to the cavity. We show that the number of QDs coupled to the cavity can be controlled by the size of the window. Furthermore, we demonstrate Purcell enhancement and saturable photoluminescence in this QD-cavity platform. Finally, we deterministically position QDs on a photonic molecule and observe QD-coupled cavity super-modes. Our results pave the way for controlling the number of QDs coupled to a cavity by engineering the window size, and the QD dimension, and will allow advanced studies in cavity enhanced single photon emission, ultralow power nonlinear optics, and quantum many-body simulations with interacting photons.

**Introduction**

Hybrid photonic integrated circuits, comprised of nanophotonic structures and active media, have recently seen an outpouring of diverse applications, ranging from ultralow threshold nanolasers[1–5] to quantum networks[6,7]. A key driver behind their success has been the improved engineering of the electromagnetic environment with nanoscale optical resonators, which have led to enhanced light-matter coupling and demonstrations of quantum optical effects in both the weak and the strong coupling regimes[8–10]. As a result, it has now become possible to fabricate a robust array of high quality (Q)-factor cavities on the same chip, opening a possible route to building multi-functional optical interconnects[11,12] as well as scalable, on-chip quantum simulators[13,14].

While state-of-the-art fabrication methods can yield hundreds of cavities with sub-wavelength precision, large-scale control over the positioning of multiple active media remains elusive. Extensive work has been carried out with self-assembled semiconductor quantum dots (QDs) to overcome their random positioning and inhomogeneous broadening, including seeding nucleation centers for site-controlled growth[15], but there has been no report of a deterministically coupled system of multiple QD-cavities. Beyond semiconductor QDs, several studies looked at deterministic creation of nanodots and single emitters using monolayer materials[16–18], albeit with limited success.

A promising candidate for active media in hybrid photonic integrated circuits is solution processed colloidal quantum dots (QDs)[19]. Owing to their robust synthesis and straightforward application onto most substrates, colloidal QDs have generated intense interest as a novel class of light emitting materials[20–22]. Optically pumped lasers and electrically triggered single photon sources based on colloidal QDs have recently been demonstrated[23–26]. Low threshold nanolasers and low power nonlinear optical devices have also been achieved by coupling the QDs to nanocavities[2,27–29]. The simple drop-cast and spin-coat methods that were employed to place the QDs on the cavities, however, are probabilistic in nature, where the only control that the experimenter has is the QD density in the solution. An innovative method to trap the colloidal QDs in lithographically defined windows has recently been demonstrated[30]. When a substrate with nanoscale windows is spin-coated with a uniform thin film of the solution, the QDs, depending on their relative size and the chemical properties of the solution, enter and occupy the windows, thus dramatically increasing the selective placement probability. Combining this method with cavities, on the other hand, is complicated as typical ultra-low mode-volume photonic crystal cavities (PhC) cannot be cladded in a thick resist without severely degrading their Q factors, since the sharpness of the cavity resonance comes from the contrast in the index of refraction between the cavity material and the surrounding environment[31]. Although it is possible to develop methods to post-liftoff the resist, it is nontrivial to guarantee that the QDs would still attach with the cavities after the processing. Moreover, the suspended nature of most PhC cavities will prevent further sonication for resist cleaning. These problems can, however, be solved using a recently demonstrated encapsulated nanobeam cavity in the silicon nitride platform[32].

In this paper, we experimentally demonstrate deterministic positioning of solution processed colloidal QDs on silicon nitride nanobeam cavities. The cavities follow the previously reported encapsulated design with elliptical holes[32], which allows them to maintain high Q-factors despite being cladded in a thick layer of Poly-methyl methacrylate (PMMA) resist. After lithographically opening up fixed-sized windows in the resist, we spin-coat the chip with a uniform film of the colloidal solution, which yields an array of coupled QD-cavities. We further verify the coupling by observing Purcell enhancement and saturable photoluminescence. Finally, we demonstrate coupling between the QDs and a pair of coupled nanobeam cavities, called a photonic molecule. Our work paves the way to creating a large array of coupled cavities with each cavity containing a specified number of QDs, with potential applications in nonlinear optics, multi-functional optical devices, and on-chip, solid-state quantum simulators.

**Encapsulated silicon nitride nanobeam cavity**

We designed, fabricated, and tested the SiN nanobeam cavity following the same process as our previously reported method[32]. We first calculated the band structure of the unit cell (MIT Photonic Bands) and optimized the whole cavity structure with finite difference time domain simulation (Lumerical). Specifically, we created the cavity by linearly tapering the major axis diameter of the holes and the period about the cavity center. We adapted 10 elliptical holes for tapering region and optimized the design parameters until we found a suitably high Q-factor (Q $\sim 10^5$) resonance centered at 630 nm. In the final design, the nanobeam has a thickness t = 220 nm and a width w = 553 nm. The Bragg region consists of 40 elliptical holes placed at a periodicity of a = 189 nm. The elliptical holes have a major and a minor diameter of 242 nm and 99 nm, respectively. In the tapering region, the periodicity and the major diameter of the hole is linearly reduced to 179 nm and 112 nm, while the minor diameter is fixed. The cavity length is 72 nm. The resulting electromagnetic field has a mode volume of ~$2.5 \left(\frac{\lambda}{n}\right)^3$, on the same order as that of a previously reported floating SiN nanobeam resonator[31].

We fabricated the cavity using 220 nm thick SiN membrane grown via LPCVD on 4 μm of thermal oxide on silicon. The samples were obtained from commercial vendor Rogue Valley Microelectronics. We spun roughly 400 nm of Zeon ZEP520A, which was coated with a thin layer of Pt/Au that served as a charging layer. The resist was then patterned using a JEOL JBX6300FX electron beam lithography system with an accelerating voltage of 100 kV. The pattern was transferred to the SiN using a RIE etch in $CHF_3/O_2$. Figures 1 a,b show the scanning electron micrographs (SEMs) of the fabricated SiN cavities on thermal oxide just after etching. Figure 1c shows the simulated profile of the mode confined in the cavity. To encapsulate the cavities, we spun ~ 1 μm PMMA at 3 krpm speed and then baked the chip to remove any remaining solvent.

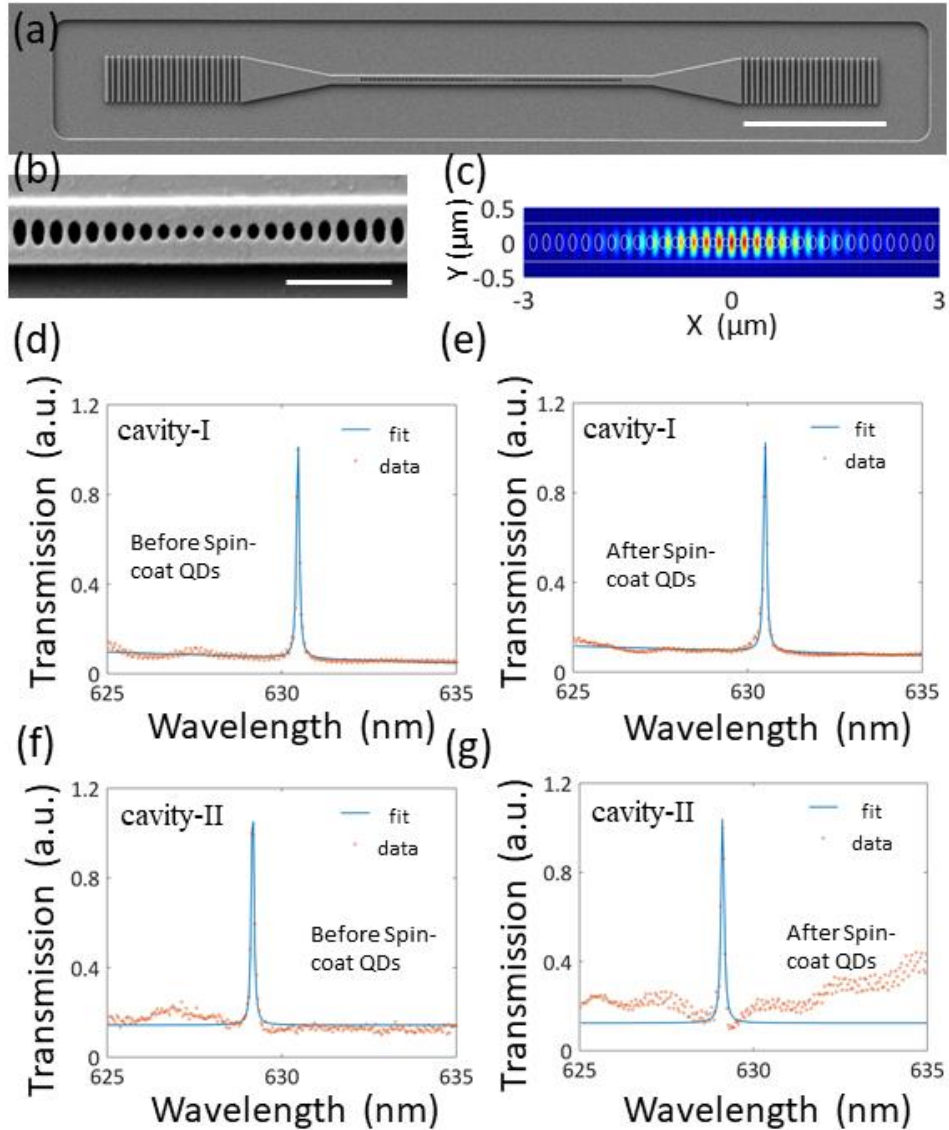

*Figure 1: Cavity transmission characterization. (a), (b) SEM of the silicon nitride cavity, where the nanobeam is unsuspended and sitting on the silicon oxide. Scale bar: 10 µm, 1 µm. (c) Simulated cavity mode profile via FDTD (d) Transmission spectrum of the cavity without a PMMA window (Cavity-I) before spin-coating colloidal QDs (Q~6900) and (e) after spin-coating colloidal QDs (Q~6600) (f) Transmission spectrum of the cavity with a PMMA window (Cavity-II) before spin-coating colloidal QDs (Q~7600) and (g) after spin-coating colloidal QDs (Q~6200). The results indicate that the cavity-I could still retain high-Q operation under organic polymer cladding. Also, due to the limited QD absorption, the spin-coating of QDs does not dramatically degrade the Q factor of Cavity-II.*

We then measured the transmission spectra of the cavities using a supercontinuum light-source (Fianium WhiteLase Micro). The supercontinuum light was focused on one of the two grating couplers, and the transmitted light collected from the other was analyzed with a spectrometer (Princeton Instruments PIXIS CCD with an IsoPlane SCT-320 Imaging Spectrograph). The use of the grating couplers to measure the cavity transmission and to collect the coupled PL of QDs in the following experiments is beneficial for on-chip light sources to be integrated with other integrated optics components. The cavity transmission spectrum is shown in Figure 1d. We clearly observed a cavity resonance at 630 nm with quality factor ~ 6600, extracted via a Lorentzian fit to the measured data (Figure 1d). We note that the experimental Q-factor is significantly smaller than our simulation, which we attribute to fabrication imperfections due to the small feature size for visible wavelength operation.

**Deterministic positioning of colloidal QDs on a single cavity**

Colloidal CdSe/CdS core-shell QDs were synthesized to have PL emission centered at 630 nm, matching the cavity resonance. The QD synthesis method and the PL spectrum of the as-prepared QDs are described in the Supplementary Materials. We first performed an overlay process using electron-beam lithography to define small square-shaped windows with different side lengths (1.5 μm, 750 nm, 500 nm, and 300 nm) in the PMMA resist that had been placed on top of the chip containing multiple nanocavities. The locations of the windows were chosen to coincide with those of the antinodes of the cavity modes. We also left some cavities inaccessible to the QDs without any PMMA window.

Following this setup, we dissolved 10 nM QD in 10:1 hexane and octane, filtered through a 450 nm Polyvinylidene Fluoride (PVDF) filter, and then spun coat the QD solution to get a uniform thin film on top of the device. From ellipsometry measurements, the QD thin film had a thickness of 80 nm and refractive index ~ 1.5. We note that while pure CdSe has a refractive index of ~ 2.3, the whole thin film has a lower index due to the presence of organic ligands and solution residues. Figure 2 shows the schematic of our device. For a cavity without a window (cavity-I), we expect to observe no coupling with the QDs, as the thick resist will prevent any coupling between the cavity and the QD layer. For cavities with windows (cavity-II and III), we expect to observe coupling with QDs and qualitatively control the coupling by varying the size of the window.

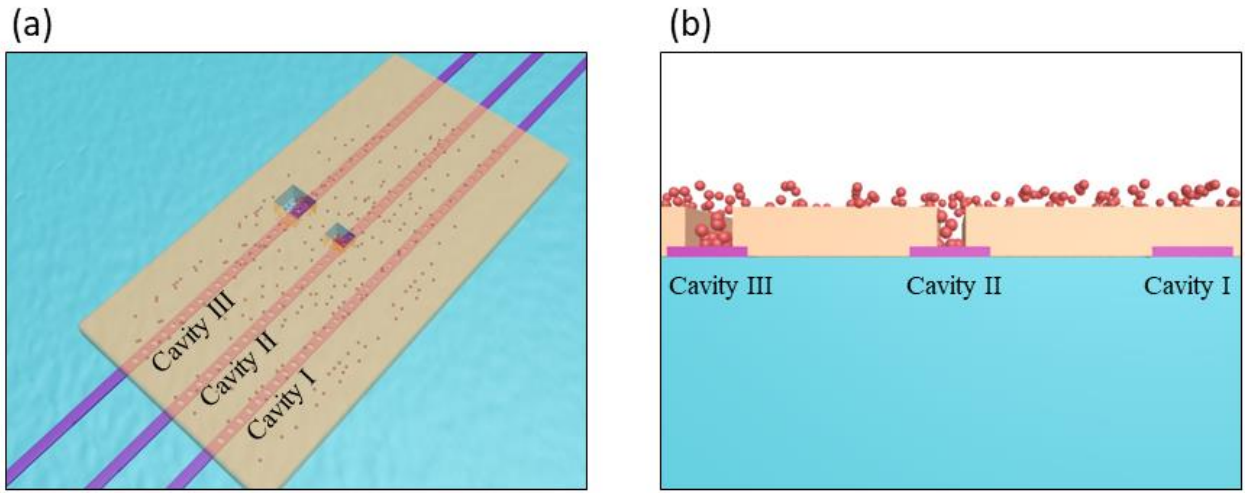

*Figure 2: Schematics of the deterministic positioning mechanism. (a) Multiple SiN nanobeam cavities (Cavity I, II, III) could be integrated on the same chip. These cavities maintain high-Q operation even under organic resist cladding. To deterministically position QDs, we selectively opened up windows on certain areas on the chip. This is followed with spin-coating QDs where the QDs filled into the windows to interact with cavities. For cavity-I, we expect to observe no coupling with the QDs, as the thick resist will prevent any coupling between the cavity and the QD layer. For cavities with windows (cavity-II and III), we expect to observe coupling with QDs and qualitatively control the coupling by varying the size of the window. (b) The cross section of the*

*Cavity I, II, III showing how the QDs enter the windows (Cavity II, III) and couple to the cavity fields.*

We first compared the device performance before and after the solution deposition. For cavities without PMMA windows, the Q factor remained the same both before and after the QD deposition, indicating that the QDs did not couple to the cavities. Figure 1e is the transmission measurement result after solution deposition. For cavities with PMMA windows, the spectrum before the electron beam exposure and solution deposition is shown in Figure 1, with Q factor of 7600. The cavity resonance disappeared after the electron beam exposure and before the solution deposition, since the change of the refractive index in the window region (filled with air) dramatically perturbed the mode and degraded the quality factor, which we confirmed by FDTD simulation. In the FDTD simulation, a cavity with quality factor of $\sim 10^5$ dropped to 1200 when a 1.5μm × 1.5μm window is opened up in its PMMA. However, after the QD deposition as shown in Figure 1g, the cavity recovered to an experimentally verified Q factor of 6200.

Having confirmed the robustness of the cavity resonance in the presence of PMMA windows, we performed the photoluminescence (PL) measurement. Figure 3a shows the SEM of the device with schematic of $1.5 \mu m$ PMMA window. Figure S2 in the supplement shows the experimental setup for the PL characterization. A continuous wave (CW) green diode laser ($\lambda \sim 532$ nm) was used to pump the center of the cavity where the PMMA window was located. The laser was focused to a 1-μm-diameter beam spot by an objective lens with NA= 0.65. We also used a 550 nm low-pass filter to block the pumping light in the collection path. We first confirmed the QD-cavity coupling by pumping the QDs and observing PL coming out of the grating couplers with a CCD camera (Figure 3b). For more detailed analysis of the light, we used a spectrometer. Since the PL signal coming from the window location was much brighter than that coming from the grating couplers, we used a pin-hole to collect the light only from the grating coupler when we were studying the cavity signal (Figure 3c). The cavity mode at 629 nm matched with our transmission measurement. We note that another mode at 612 nm appeared in the PL measurement compared to just a single mode observed for the cavity before the QDs were applied. We attribute this to the slight refractive index difference of the QDs with PMMA. The higher refractive index of the QDs breaks the z-directional symmetry of the cavity, and through numerical simulation, we confirmed it was indeed a new TM mode[32] (see Supplementary Materials). However, as shown in Fig. 3c, for a cavity with no PMMA window, when we collected PL signal from the grating we only observe scattered background signal and no cavity signal. We were able to observe coupling down to the smallest window (300 nm side length) on the chip, indicating our deterministic positioning mechanism is robust. Further improvement of the viscosity of the solution should allow the QDs to get into even smaller windows. In addition to tuning the spatial position for controllable coupling, we also achieved spectral control of the PL coupled to the cavity by fabricating cavities with linear change of Bragg period on the same chip. Figure 4a shows the PL coupled with cavities with different resonance, covering the whole PL emission spectral region of the QDs.

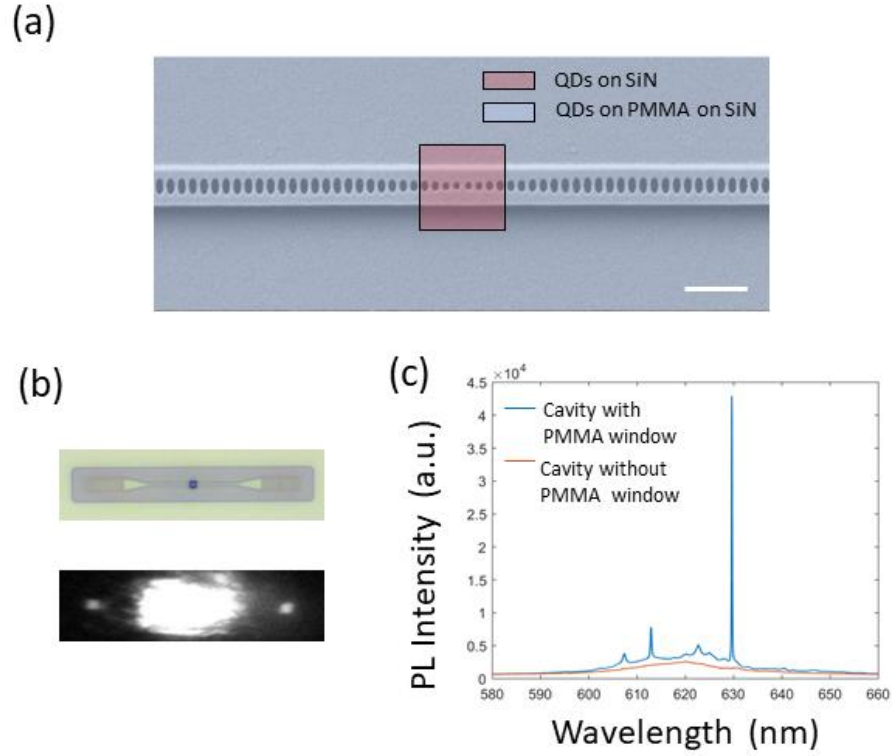

*Figure 3: PL characterization of the QD-cavity. (a) SEM of the cavity-II. Scale bar: 1.5μm. A schematic of the outline of the opened window is superimposed with the SEM. (b) An optical microscope image showing the opening on the cavity. The image of the cavity captured in the PL measurement setup after pumping the cavity-II. The lighting up of the grating couplers indicates the coupling between the QDs and cavity. (c) PL spectrum: For a cavity with a PMMA window, the cavity signal(629nm,612nm) is clearly observed against the PL background. A new TM mode at 612nm appears compared with the transmission measurement, originating from the slightly higher refractive index of the QDs breaking the z-directional symmetry of the cavity. For a cavity without a PMMA window, no cavity coupling is observed, as expected.*

We further confirmed the cavity enhancement by performing lifetime measurements (Figure 4b). We fitted the data with a multiexponential decay model[33]:

$$I(t) = I_0 + Ae^{-(t/\gamma_0)^\beta}$$

The average lifetime is given by:

$$\gamma_{avg} = \frac{\gamma_0}{\beta} \Gamma\left(\frac{1}{\beta}\right)$$

The Purcell enhancement factor is given by:

$$F_{max} = 1 + \frac{3\lambda^3}{4\pi^2 n^2} \frac{Q_{np}}{V} \psi(r)$$

Here, $\lambda$ is the cavity resonance wavelength; $Q_{np}$ is the quality factor of the quantum dot emission linewidth; $n$ is the refractive index of the cavity dielectric; $V$ is the cavity mode volume and $\psi(r)$ is the ratio of mode intensity at the emitter's location to the maximum. We note that we are using the quality factor of the emitter but not the cavity since we are in the "bad" emitter regime, where the linewidth of the emitter is much larger than the linewidth of the cavity[34]. For our device, the linewidth of the QD emission was 23 nm, giving a quality factor of 27; the numerically estimated mode volume is $2.5 \left(\frac{\lambda}{n}\right)^3$; $\psi(r)$ is 0.35 as the QD interacts only with the evanescent field of the cavity; the refractive index of SiN is 2. With these values, the theoretically calculated Purcell factor is 1.4. We extracted a lifetime of 4.8 ns for the PL emission and 3.8 ns for the cavity coupled PL emission, indicating a Purcell factor of 1.26. The slight discrepancy between the measured Purcell enhancement and the theory is attributed to the fact that some of the QDs were not located at the field maximum on the surface. We note that due to our higher mode volume compared to suspended cavities, our Purcell enhancement factor was smaller than the largest value (4.2) reported in a dielectric resonator[27]. However, by further optimization, a lower mode-volume resonator can be realized[35]. For example, by exploring a nanobeam design with a slot structure[36], one could dramatically reduce the mode volume while maintaining a high Q factor, and thus a much higher enhancement factor.

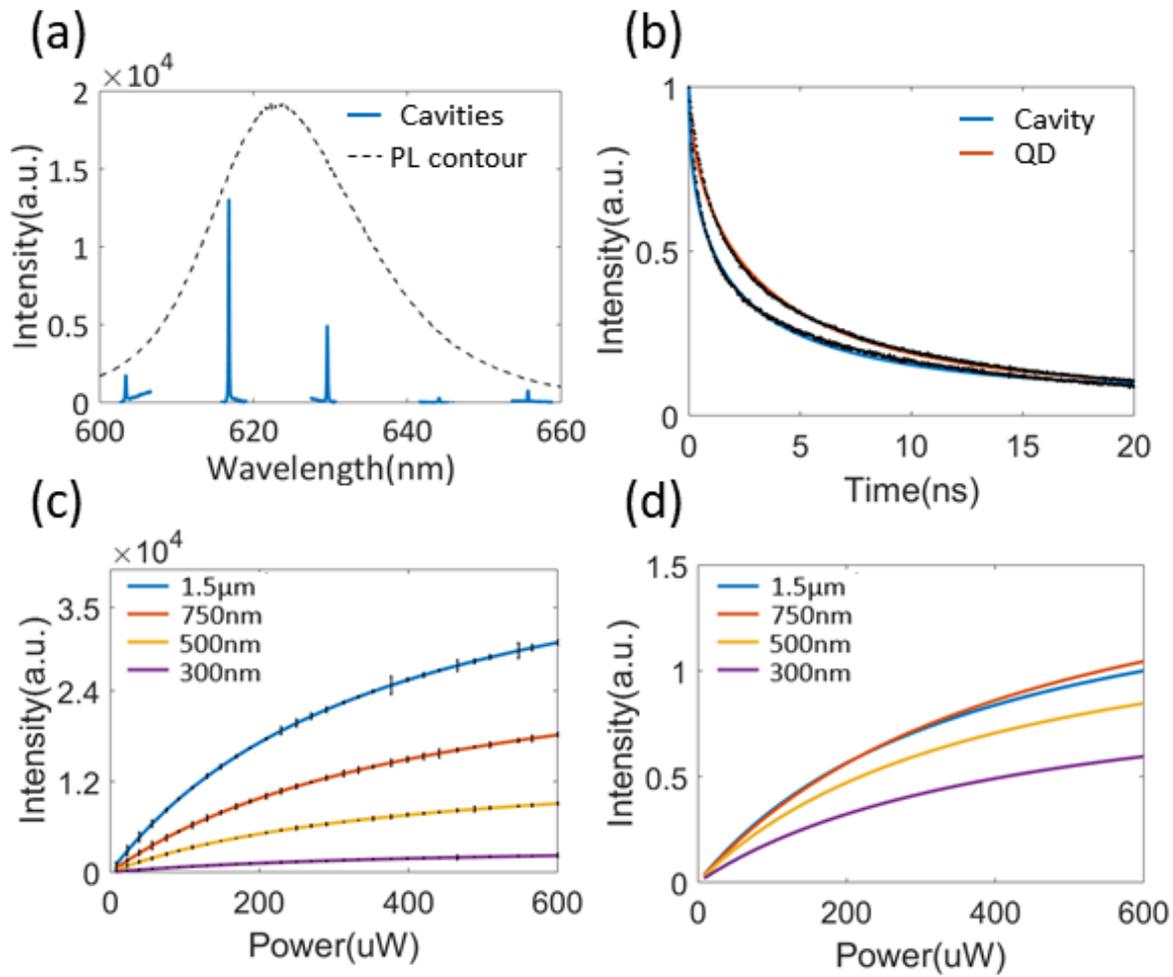

*Figure 4: Spectral and spatial control of the QD-cavity coupling: (a) We show the cavity-coupled PL over the whole resonance spectrum by positioning QDs on cavities with scaling geometry. The black dotted curve shows the contour of the PL. (b) Lifetime measurement: The solid red and blue curves are the fits to the time-resolved PL signal from the QDs on substrate and the QDs coupled with the cavity, respectively. The black dots are the raw experimental data. A Purcell factor of 1.26 is measured. (c) Power series for cavities with PMMA windows with different sizes: 1.5μm × 1.5μm, 750nm×750nm, 500nm×500nm, 300nm×300nm. As the size of the window grows, the cavity signal in PL increases since more QDs are interacting with the cavity (d) Power series for cavity-coupled PL normalized by the mode area of the cavity inside the window region.*

To further explore the possibility of controlling the number of QDs coupled to the cavities, we performed power series measurements of samples with different window sizes. The difference in the photoluminescence intensity was observed; cavities with larger windows had brighter emission in general. To get a more quantitative understanding of how the size of the window affected the number of QDs coupling with the cavity, we normalized the emission intensity according to the cavity mode area exposed by the windows. From the FDTD simulation, the mode areas for the $1.5\ \mu m$, $760\ nm$, $500\ nm$, and $300\ nm$ window are, $0.23 \mu m^2$, $0.13 \mu m^2$, $0.08 \mu m^2$, $0.03\ \mu m^2$, respectively. We saw that the intensity curves for the 1550 nm and 750 nm windows almost overlapped on top of each other after the normalization. For the device with 500 nm and 300 nm windows, however, the intensities were lower than those with the larger window cavities, with the intensity for the 300 nm window even lower than that for 500 nm window. We attribute this observation to the fact that as the windows become smaller, the QDs are no longer able to enter into the cavities efficient due to the surface tension of the solution. However, further surface modification and solution with lower viscosity could potentially allow more QDs to enter through the windows. For all the window sizes examined, we observed that the photoluminescence saturated when pumped with increasing laser power. We fitted the data and extracted saturation power ~400 $\mu$W (see supplement S4). We did not observe significant difference in the saturation power for different window sizes, since the intensity of the pumping light on each QD was essentially the same in all four cases.

**Coupling of QDs to a photonic molecule**

One promising application of our deterministic positioning method is performing quantum many-body simulations[37] using QDs coupled to cavity arrays. The simplest array, made up of just a pair of coupled cavities, is called a photonic molecule[38]. It has been shown in several theoretical studies that QDs coupled to a photonic molecule may form the basis for studying exotic phases of matter[39] and other cavity quantum electrodynamics phenomena such as unconventional photon blockade[38,40]. However, both scalability and deterministic positioning are difficult to achieve with conventional self-assembled semiconductor QDs coupled with suspended coupled nanobeam cavities. Besides, the mode symmetric nature of the coupled cavity super-modes also precludes the reflection measurement of photonic crystals by directly pumping and collecting a laser signal at the center of the cavity[41]. Here we fabricate the photonic molecule with grating couplers for each cavity for transmission measurements and deterministically position the QDs to couple with the cavity super modes.

Figure 5a shows the SEM of the fabricated device. Each cavity has a pair of grating couplers that allows measuring transmission from each cavity independently. We fabricate two coupled cavities with different gaps between them: 1.5 μm, 400 nm, and 200 nm (Figure 5b). Figure 5c shows the transmission spectrum measured via the grating for cavity 1. For cavities 1.5 μm apart, we see only one cavity in transmission, indicating there is no coupling between two cavities. For cavities 400 nm and 200 nm apart, we observe the two coupled super-modes. As the distance becomes smaller for the two cavities, the coupling strength becomes stronger, resulting in larger spectral separation of the two modes.

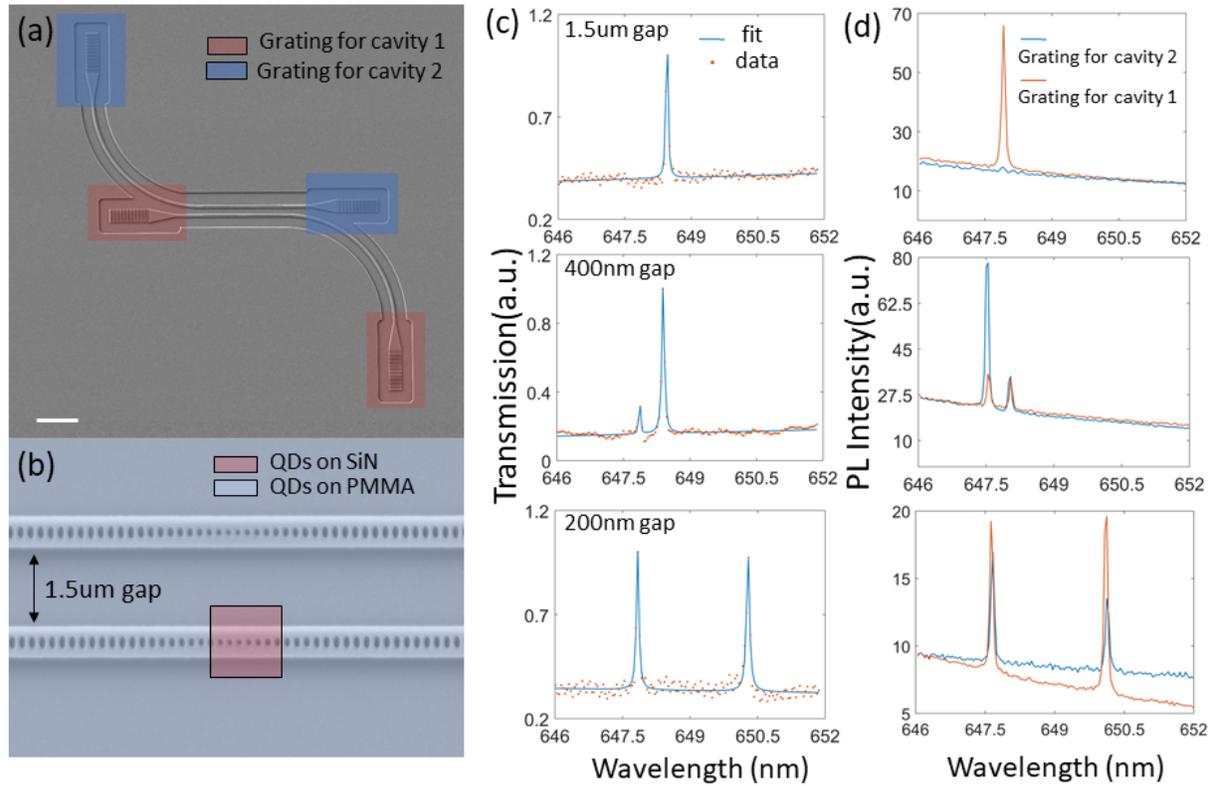

*Figure 5: Deterministic positioning of QDs to a photonic molecule. (a) SEM image of the photonic molecule. Each cavity has a pair of grating couplers for collecting and extracting the QDs' PL. Scale bar: 10μm. (b) Schematic of the outline of the opened window superimposed with the SEM of the device. (c)Transmission measurement of the device with different intra-cavity gaps before spin-coating QDs. For cavities 1.5μm apart, we saw only one cavity resonance in transmission, indicating no coupling between the two cavities. For cavities 400nm and 200nm apart, as the distance becomes smaller for the two cavities, the coupling strength becomes stronger, resulting in larger spectral separation of the two supermodes. (d) PL characterization: For cavities 1.5um apart, we observed the cavity signal from the grating for cavity 2, since the PL signal was only coupled with cavity 1 and the two cavities were not coupled with each other. For cavities 400nm and 200nm apart, we successfully observed the coupling between the QDs and the super-modes at both gratings for cavity 1 and cavity 2.*

We then opened up PMMA windows on cavity 2 and spin-coated it with the QD solution. We adjusted the collimation of the pumping beam so that it illuminated both cavities, and we collected PL from gratings for both cavities. The results are shown in Figure 5d. For cavities that were 1.5 µm apart, we only observed the cavity signal from the grating for cavity 2, since the gap was too large for the two cavities to couple. For cavities 400 nm and 200 nm apart, we successfully observed coupling between the QDs and the super-modes at both gratings for cavity 1 and cavity 2. This approach can be readily scaled up to an array of multiple coupled QD-cavities.

**Conclusions**

In summary, by selectively opening up a PMMA window on an encapsulated SiN nanobeam cavity and performing solution-phase deposition, we have demonstrated deterministic coupling between colloidal QDs and an encapsulated silicon nitride nanobeam cavity. We have also explored the coupling between the colloidal QDs and a photonic molecule. Our results suggest several directions in future research, one of which is to tailor the size of the window as well as the QDs to create an array of coupled cavities with exactly one QD per window. This could be done by, for instance, creating giant core-shell QDs[42,43]. Our results also pave the way for future studies of colloidal QDs coupled with various photonic crystal cavity platforms, with applications in cavity enhanced single photon emission, low power nonlinear optics, and quantum many body simulations.

**Acknowledgement**

We thank Michael Enright for helpful discussions in the preparation of the CdSe quantum dots. This work is supported by the National Science Foundation under grant NSF-EFRI-1433496, NSF-ECCS-1708579, NSF MRSEC 1719797, the Air Force Office of Scientific Research grant FA9550-18-1-0104, Alfred P. Sloan research fellowship and the David and Lucile Packard Foundation. Max Friedfeld is a Washington Research Foundation Postdoctoral Fellow. All the fabrication processes were performed at the Washington Nanofabrication Facility (WNF), a National Nanotechnology Infrastructure Network (NNIN) site at the University of Washington, which is supported in part by the National Science Foundation (awards 0335765 and 1337840), the Washington Research Foundation, the M. J. Murdock Charitable Trust, GCE Market, Class One Technologies, and Google.

**REFERENCES**

(1) Wu, S.; Buckley, S.; Schaibley, J. R.; Feng, L.; Yan, J.; Mandrus, D. G.; Hatami, F.; Yao, W.; Vučković, J.; Majumdar, A.; et al. Monolayer Semiconductor Nanocavity Lasers with Ultralow Thresholds. *Nature* **2015**, *520* (7545), 69–72.
(2) Ellis, B.; Mayer, M. A.; Shambat, G.; Sarmiento, T.; Harris, J.; Haller, E. E.; Vučković, J. Ultralow-Threshold Electrically Pumped Quantum-Dot Photonic-Crystal Nanocavity Laser. *Nature Photonics* **2011**, *5* (5), 297–300.
(3) Dai, D.; Fang, A.; Bowers, J. E. Hybrid Silicon Lasers for Optical Interconnects. *New J. Phys.* **2009**, *11* (12), 125016.
(4) Li, Y.; Zhang, J.; Huang, D.; Sun, H.; Fan, F.; Feng, J.; Wang, Z.; Ning, C. Z. Room-Temperature Continuous-Wave Lasing from Monolayer Molybdenum Ditelluride Integrated with a Silicon Nanobeam Cavity. *Nature Nanotechnology* **2017**, *12* (10), 987–992.
(5) Jagsch, S. T.; Triviño, N. V.; Lohof, F.; Callsen, G.; Kalinowski, S.; Rousseau, I. M.; Barzel, R.; Carlin, J.-F.; Jahnke, F.; Butté, R.; et al. A Quantum Optical Study of Thresholdless Lasing Features in High- β Nitride Nanobeam Cavities. *Nature Communications* **2018**, *9* (1), 564.
(6) Faraon, A.; Majumdar, A.; Englund, D.; Kim, E.; Bajcsy, M.; Vučković, J. Integrated Quantum Optical Networks Based on Quantum Dots and Photonic Crystals. *New J. Phys.* **2011**, *13* (5), 055025.
(7) Davanco, M.; Liu, J.; Sapienza, L.; Zhang, C.-Z.; Cardoso, J. V. D. M.; Verma, V.; Mirin, R.; Nam, S. W.; Liu, L.; Srinivasan, K. Heterogeneous Integration for On-Chip Quantum

(7) Photonic Circuits with Single Quantum Dot Devices. *Nature Communications* **2017**, *8* (1), 889.
(8) Buckley, S.; Rivoire, K.; Vučković, J. Engineered Quantum Dot Single-Photon Sources. *Rep. Prog. Phys.* **2012**, *75* (12), 126503.
(9) Englund, D.; Majumdar, A.; Faraon, A.; Toishi, M.; Stoltz, N.; Petroff, P.; Vučković, J. Resonant Excitation of a Quantum Dot Strongly Coupled to a Photonic Crystal Nanocavity. *Phys. Rev. Lett.* **2010**, *104* (7), 073904.
(10) Reinhard, A.; Volz, T.; Winger, M.; Badolato, A.; Hennessy, K. J.; Hu, E. L.; Imamoğlu, A. Strongly Correlated Photons on a Chip. *Nature Photonics* **2012**, *6* (2), 93–96.
(11) Liang, D.; Huang, X.; Kurczveil, G.; Fiorentino, M.; Beausoleil, R. G. Integrated Finely Tunable Microring Laser on Silicon. *Nature Photonics* **2016**, *10* (11), 719–722.
(12) Heck, M. J. R.; Bauters, J. F.; Davenport, M. L.; Doylend, J. K.; Jain, S.; Kurczveil, G.; Srinivasan, S.; Tang, Y.; Bowers, J. E. Hybrid Silicon Photonic Integrated Circuit Technology. *IEEE Journal of Selected Topics in Quantum Electronics* **2013**, *19* (4), 6100117–6100117.
(13) Georgescu, I. M.; Ashhab, S.; Nori, F. Quantum Simulation. *Rev. Mod. Phys.* **2014**, *86* (1), 153–185.
(14) Aspuru-Guzik, A.; Walther, P. Photonic Quantum Simulators. *Nature Physics* **2012**, *8* (4), 285–291.
(15) Schneider, C.; Strauß, M.; Sünner, T.; Huggenberger, A.; Wiener, D.; Reitzenstein, S.; Kamp, M.; Höfling, S.; Forchel, A. Lithographic Alignment to Site-Controlled Quantum Dots for Device Integration. *Appl. Phys. Lett.* **2008**, *92* (18), 183101.
(16) Wei, G.; Czaplewski, D. A.; Lenferink, E. J.; Stanev, T. K.; Jung, I. W.; Stern, N. P. Size-Tunable Lateral Confinement in Monolayer Semiconductors. *Scientific Reports* **2017**, *7* (1), 3324.
(17) Ryou, A.; Rosser, D.; Saxena, A.; Fryett, T.; Majumdar, A. Strong Photon Antibunching in Weakly Nonlinear Two-Dimensional Exciton-Polaritons. *Phys. Rev. B* **2018**, *97* (23), 235307.
(18) Palacios-Berraquero, C.; Kara, D. M.; Montblanch, A. R.-P.; Barbone, M.; Latawiec, P.; Yoon, D.; Ott, A. K.; Loncar, M.; Ferrari, A. C.; Atatüre, M. Large-Scale Quantum-Emitter Arrays in Atomically Thin Semiconductors. *Nature Communications* **2017**, *8*, 15093.
(19) Alivisatos, A. P. Semiconductor Clusters, Nanocrystals, and Quantum Dots. *Science* **1996**, *271* (5251), 933–937.
(20) Qu, L.; Peng, X. Control of Photoluminescence Properties of CdSe Nanocrystals in Growth. *J. Am. Chem. Soc.* **2002**, *124* (9), 2049–2055.
(21) Anikeeva, P. O.; Halpert, J. E.; Bawendi, M. G.; Bulović, V. Quantum Dot Light-Emitting Devices with Electroluminescence Tunable over the Entire Visible Spectrum. *Nano Lett.* **2009**, *9* (7), 2532–2536.
(22) Chandrasekaran, V.; Tessier, M. D.; Dupont, D.; Geiregat, P.; Hens, Z.; Brainis, E. Nearly Blinking-Free, High-Purity Single-Photon Emission by Colloidal InP/ZnSe Quantum Dots. *Nano Lett.* **2017**, *17* (10), 6104–6109.
(23) le Feber, B.; Prins, F.; De Leo, E.; Rabouw, F. T.; Norris, D. J. Colloidal-Quantum-Dot Ring Lasers with Active Color Control. *Nano Lett.* **2018**, *18* (2), 1028–1034.
(24) Guzelturk, B.; Kelestemur, Y.; Gungor, K.; Yeltik, A.; Akgul, M. Z.; Wang, Y.; Chen, R.; Dang, C.; Sun, H.; Demir, H. V. Stable and Low-Threshold Optical Gain in CdSe/CdS

(24) Quantum Dots: An All-Colloidal Frequency Up-Converted Laser. *Advanced Materials* **27** (17), 2741–2746.
(25) Fan, F.; Voznyy, O.; Sabatini, R. P.; Bicanic, K. T.; Adachi, M. M.; McBride, J. R.; Reid, K. R.; Park, Y.-S.; Li, X.; Jain, A.; et al. Continuous-Wave Lasing in Colloidal Quantum Dot Solids Enabled by Facet-Selective Epitaxy. *Nature* **2017**, *544* (7648), 75–79.
(26) Lin, X.; Dai, X.; Pu, C.; Deng, Y.; Niu, Y.; Tong, L.; Fang, W.; Jin, Y.; Peng, X. Electrically-Driven Single-Photon Sources Based on Colloidal Quantum Dots with near-Optimal Antibunching at Room Temperature. *Nature Communications* **2017**, *8* (1), 1132.
(27) Gupta, S.; Waks, E. Spontaneous Emission Enhancement and Saturable Absorption of Colloidal Quantum Dots Coupled to Photonic Crystal Cavity. *Opt. Express, OE* **2013**, *21* (24), 29612–29619.
(28) Faraon, A.; Fushman, I.; Englund, D.; Stoltz, N.; Petroff, P.; Vučković, J. Coherent Generation of Non-Classical Light on a Chip via Photon-Induced Tunnelling and Blockade. *Nature Physics* **2008**, *4* (11), 859–863.
(29) Yang, Z.; Pelton, M.; Fedin, I.; Talapin, D. V.; Waks, E. A Room Temperature Continuous-Wave Nanolaser Using Colloidal Quantum Wells. *Nature Communications* **2017**, *8* (1), 143.
(30) Xie, W.; Gomes, R.; Aubert, T.; Bisschop, S.; Zhu, Y.; Hens, Z.; Brainis, E.; Van Thourhout, D. Nanoscale and Single-Dot Patterning of Colloidal Quantum Dots. *Nano Lett.* **2015**, *15* (11), 7481–7487.
(31) Khan, M.; Babinec, T.; McCutcheon, M. W.; Deotare, P.; Lončar, M. Fabrication and Characterization of High-Quality-Factor Silicon Nitride Nanobeam Cavities. *Opt. Lett., OL* **2011**, *36* (3), 421–423.
(32) Fryett, T. K.; Chen, Y.; Whitehead, J.; Peycke, Z. M.; Xu, X.; Majumdar, A. Encapsulated Silicon Nitride Nanobeam Cavity for Hybrid Nanophotonics. *ACS Photonics* **2018**.
(33) van Driel, A. F.; Nikolaev, I. S.; Vergeer, P.; Lodahl, P.; Vanmaekelbergh, D.; Vos, W. L. Statistical Analysis of Time-Resolved Emission from Ensembles of Semiconductor Quantum Dots: Interpretation of Exponential Decay Models. *Phys. Rev. B* **2007**, *75* (3), 035329.
(34) Scully, M. O.; Zubairy, M. S. Quantum Optics by Marlan O. Scully /core/books/quantum-optics/08DC53888452CBC6CDC0FD8A1A1A4DD7 (accessed Jun 17, 2018).
(35) Ryckman, J. D.; Weiss, S. M. Low Mode Volume Slotted Photonic Crystal Single Nanobeam Cavity. *Appl. Phys. Lett.* **2012**, *101* (7), 071104.
(36) Choi, H.; Heuck, M.; Englund, D. Self-Similar Nanocavity Design with Ultrasmall Mode Volume for Single-Photon Nonlinearities. *Phys. Rev. Lett.* **2017**, *118* (22), 223605.
(37) Carusotto, I.; Ciuti, C. Quantum Fluids of Light. *Rev. Mod. Phys.* **2013**, *85* (1), 299–366.
(38) Majumdar, A.; Rundquist, A.; Bajcsy, M.; Vučković, J. Cavity Quantum Electrodynamics with a Single Quantum Dot Coupled to a Photonic Molecule. *Phys. Rev. B* **2012**, *86* (4), 045315.
(39) Grujic, T.; Clark, S. R.; Jaksch, D.; Angelakis, D. G. Non-Equilibrium Many-Body Effects in Driven Nonlinear Resonator Arrays. *New J. Phys.* **2012**, *14* (10), 103025.
(40) Flayac, H.; Savona, V. Unconventional Photon Blockade. *Phys. Rev. A* **2017**, *96* (5), 053810.
(41) Deotare, P. B.; McCutcheon, M. W.; Frank, I. W.; Khan, M.; Lončar, M. Coupled Photonic Crystal Nanobeam Cavities. *Appl. Phys. Lett.* **2009**, *95* (3), 031102.

(42) Chen, Y.; Vela, J.; Htoon, H.; Casson, J. L.; Werder, D. J.; Bussian, D. A.; Klimov, V. I.; Hollingsworth, J. A. "Giant" Multishell CdSe Nanocrystal Quantum Dots with Suppressed Blinking. *J. Am. Chem. Soc.* **2008**, *130* (15), 5026–5027.

(43) Kundu, J.; Ghosh, Y.; Dennis, A. M.; Htoon, H.; Hollingsworth, J. A. Giant Nanocrystal Quantum Dots: Stable Down-Conversion Phosphors That Exploit a Large Stokes Shift and Efficient Shell-to-Core Energy Relaxation. *Nano Lett.* **2012**, *12* (6), 3031–3037.

# Supplementary material: Deterministic positioning of colloidal quantum dots on silicon nitride nanobeam cavities

*Yueyang Chen[1], Albert Ryou[1], Max Friedfeld[3], Taylor Fryett[1], James Whitehead[1], Brandi Cossairt[3], Arka Majumdar[1,2]*

[1] *Electrical Engineering, University of Washington, Seattle, WA 98189, USA*

[2] *Department of Physics, University of Washington, Seattle, WA 98189, USA*

[3] *Department of Chemistry, University of Washington, Seattle, WA 98189, USA*

**S1. QD synthesis method:**

All glassware was dried in a 160 °C oven overnight prior to use. All reactions, unless otherwise noted, were run under an inert atmosphere of nitrogen using a glovebox or using standard Schlenk techniques. Cadmium oxide (99.99%), octadecylphosphonic acid (97%), ODPA, Se powder (99.99%), S (99.5%), oleic acid (OA, 90%), 1-octadecene (1-ODE, 90%), trioctylphosphine (TOP, 97%), and trioctylphosphine oxide (90%) were purchased from Sigma-Aldrich Chemical Co. and used without further purification. Solvents, including toluene, pentane, and acetonitrile, were purchased from Sigma-Aldrich Chemical Co., dried over $CaH_2$, distilled, and stored over 4 Å molecular sieves in a nitrogen-filled glovebox. Anhydrous methanol and isopropanol were purchased from Sigma-Aldrich Chemcial Co., and used as is. UV–vis spectra were collected on a Cary 5000 spectrophotometer from Agilent. Fluorescence and quantum yield measurements were taken on a Horiba Jobin Yvon FluoroMax-4 fluorescence spectrophotometer with the QuantaPhi integrating sphere accessory.

The syntheses of the wurtzite CdSe cores and the CdSe/CdS core-shell quantum dots were accomplished by following a literature procedure[1], with the following modifications. The synthesis of the CdSe cores was stopped after 30 seconds, and following work up, had the absorbance and emission (λmax = 576 nm) features shown in the figure below. In the shelling procedure, 2 mmol of CdO were used, and the amount of oleic acid was scaled accordingly. Following workup, the core-shell quantum dots had the absorbance and emission (λmax = 630 nm) features shown below.

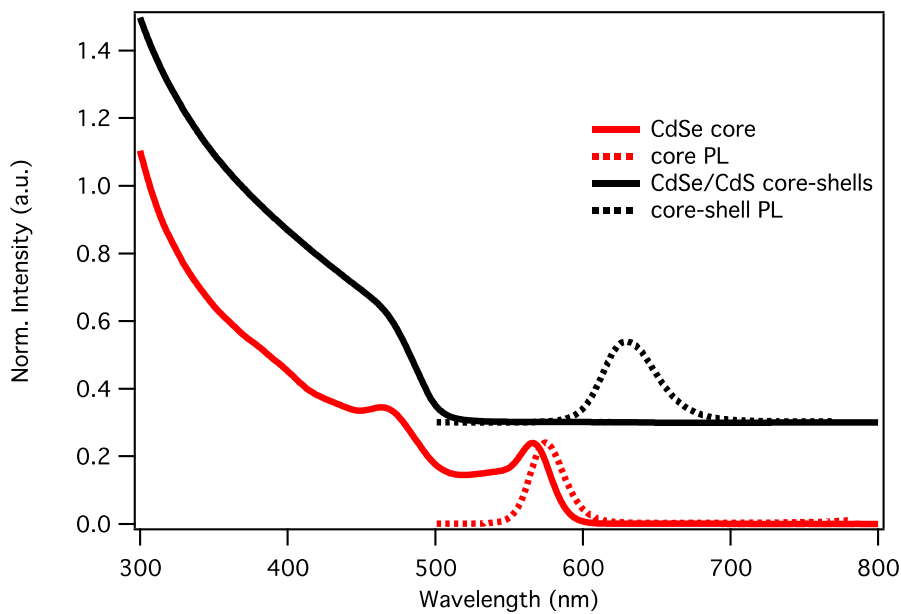

*Figure S1. UV-Vis and PL spectra of CdSe core and CdSe/CdS core-shell quantum dots. PL excitation was 400 nm.*

## S2. PL characterization setup

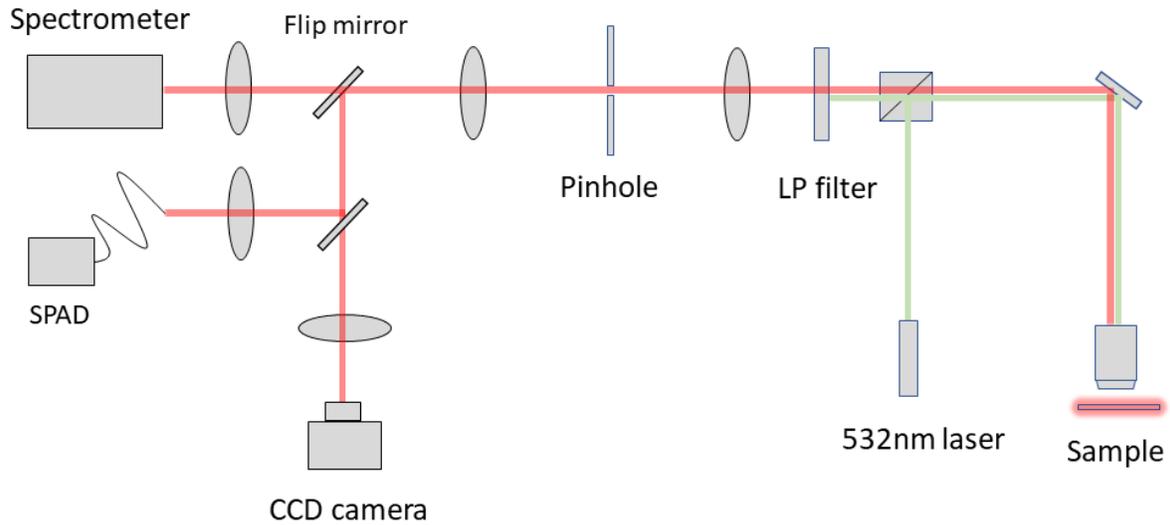

*Figure S2. Optics setup for PL characterization. A continuous wave (CW) green diode laser (λ ~ 532 nm) was used to pump the center of the cavity where the PMMA window was located. The laser was focused to a 1-µm-diameter beam spot by an objective lens with NA= 0.65. We also used a 550 nm low-pass filter to block the pumping light in the collection path. The collected light could be selectively sent into CCD camera or the spectrometer or the single-photon avalanche detector by the flip mirrors.*

S3. FDTD simulation

1. Mode profile of the TE mode and the new TM mode after spin-coating QDs. In FDTD simulation, a cladding with refractive index 1.5 is applied to the original encapsulated design structure.

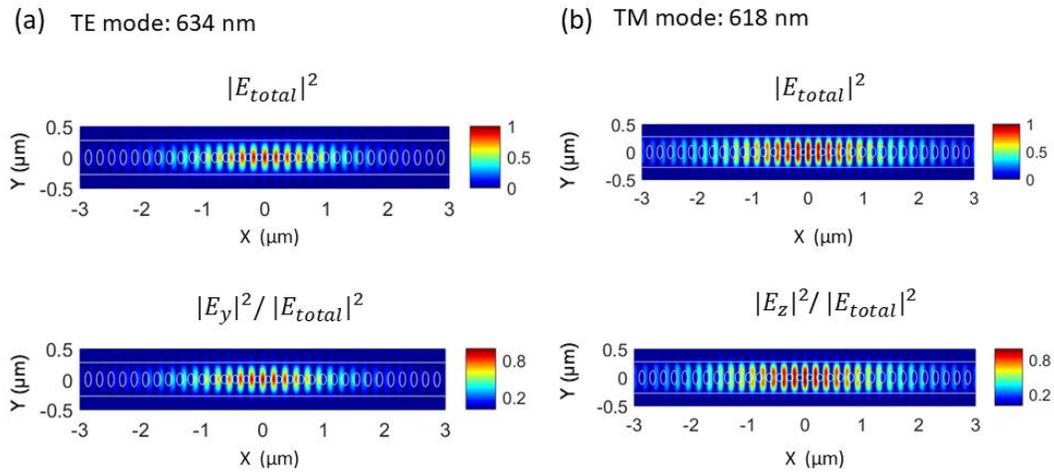

*Figure S31. Mode profile of the TE TM mode. The main electrical field components of the TE and TM mode are Ey and Ez, respectively.*

## S4. Power series fitting

We used a saturable photoluminescence model to fit the raw data from the power series measurement.

$$I_c = \frac{\alpha P}{1 + P/P_{sat}}$$

We extract the fitting parameters as follow:

| Window side length | 1.5 um | 750nm | 500nm | 300nm |
|---|---|---|---|---|
| $\alpha$ | 131 | 70 | 38 | 8 |
| $P_{sat}(\mu W)$ | 390 | 450 | 400 | 440 |

We did not observe significant difference in the saturation power for different window sizes, since the intensity of the pumping light on each QD was essentially the same in all four cases.

(1) Cirillo, M.; Aubert, T.; Gomes, R.; Van Deun, R.; Emplit, P.; Biermann, A.; Lange, H.; Thomsen, C.; Brainis, E.; Hens, Z. "Flash" Synthesis of CdSe/CdS Core–Shell Quantum Dots. *Chem. Mater.* **2014**, *26* (2), 1154–1160.